\documentclass[sigconf]{acmart}

\usepackage{amsmath,amssymb,amsfonts}
\usepackage{algorithmic}
\usepackage{graphicx}
\usepackage{textcomp}
\usepackage{xcolor}
\usepackage{soul}
\usepackage{algorithm}
\usepackage{multirow}
\usepackage{subcaption}
\usepackage{threeparttable}
\usepackage{colortbl}
\definecolor{mygray}{gray}{.85}
\usepackage{enumitem}

\def\BibTeX{{\rm B\kern-.05em{\sc i\kern-.025em b}\kern-.08em
    T\kern-.1667em\lower.7ex\hbox{E}\kern-.125emX}}

\setcopyright{acmlicensed}
\copyrightyear{2025}
\acmYear{2025}
\acmDOI{}
\acmConference[SIGSPATIAL '25]{33rd ACM SIGSPATIAL
  International Conference on Advances in Geographic Information Systems}{November 3 -- 6, 2025}{Minneapolis, MN, USA}
\acmBooktitle{33rd ACM SIGSPATIAL
International Conference on Advances in Geographic Information Systems (SIGSPATIAL '25), November 3 -- 6, 2025, Minneapolis, MN, USA}
\acmISBN{}

\begin{document}

\title[REALISM: A Regulatory Multi-Operator Shared Micromobility Vehicle Scheduling Framework]{REALISM: A Regulatory Framework for Coordinated Scheduling in Multi-Operator Shared Micromobility Services}

\author{Heng Tan}
\affiliation{%
  \institution{Lehigh University}
  \city{Bethlehem, PA}
  \country{USA}}
\email{het221@lehigh.edu}

\author{Hua Yan}
\affiliation{%
  \institution{Lehigh University}
  \city{Bethlehem, PA}
  \country{USA}}
\email{huy222@lehigh.edu}

\author{Yukun Yuan}
\affiliation{%
  \institution{University of Tennessee at Chattanooga}
  \city{Chattanooga, TN}
  \country{USA}}
\email{yukun-yuan@utc.edu}

\author{Guang Wang}
\affiliation{%
  \institution{Florida State University}
  \city{Tallahassee, FL}
  \country{USA}}
\email{guang@cs.fsu.edu}

\author{Yu Yang}
\affiliation{%
  \institution{Lehigh University}
  \city{Bethlehem, PA}
  \country{USA}}
\email{yuyang@lehigh.edu}

\ccsdesc[500]{Applied computing~Transportation}
\ccsdesc[300]{Computing methodologies~Planning and scheduling}

\keywords{Shared Micromobility; Rebalancing; Reinforcement Learning}

\begin{abstract}
Shared micromobility (e.g., shared bikes and electric scooters), as a kind of emerging urban transportation, has become more and more popular in the world. However, the blooming of shared micromobility vehicles brings some social problems to the city (e.g., overloaded vehicles on roads, and the inequity of vehicle deployment), which deviate from the city regulator's expectation of the service of the shared micromobility system. In addition, the multi-operator shared micromobility system in a city complicates the problem because of their non-cooperative self-interested pursuits. Existing regulatory frameworks of multi-operator vehicle rebalancing generally assume the intrusive control of vehicle rebalancing of all the operators, which is not practical in the real world. To address this limitation, we design REALISM, a regulatory framework for coordinated scheduling in multi-operator shared micromobility services that incorporates the city regulator's regulations in the form of assigning a score to each operator according to the city goal achievements and operators' individual contributions to achieving the city goal, measured by Shapley value. To realize the fairness-aware score assignment, we measure the fairness of assigned scores and use them as one of the components to optimize the score assignment model. To optimize the whole framework, we develop an alternating procedure to make operators and the city regulator interact with each other until convergence. We evaluate our framework based on real-world e-scooter usage data in Chicago. Our experiment results show that our method achieves a performance gain of at least 39.93$\%$ in the equity of vehicle usage and 1.82$\%$ in the average demand satisfaction of the whole city. 
\end{abstract}

\maketitle

\section{Introduction}
Shared micromobility (e.g., shared bikes and scooters) has emerged as a popular mode of urban transportation worldwide.
For instance, by 2023, 207 cities in the United States have already deployed shared micromobility systems~\cite{escooter_us}.
In place of conventional cars, shared micromobility offers a more efficient and eco-friendly mode of short trips, such as commuting from leaving subway stations for workplaces~\cite{zhong2022bike, zhong2024adatrans,zhong2023rlife}.
Service operators routinely (e.g., daily) redistribute vehicles across different areas to align supply with demand.
However, the blooming of shared micromobility vehicles introduces several problems to cities, such as overloaded vehicles on roads and spatial inequity of vehicle deployment~\cite{dias2023shared, e_scooter_problem}, which deviates from city expectations such as enhancing accessibility~\cite{escooter_guide}.
In addition, the shared micromobility system managed by multiple operators in a city (e.g., more than three operators in Chicago \cite{chicago_data}) further complicates those problems because it is challenging for cities to coordinate these operators when the operators have their own profit goals~\cite{aarhaug2023price}.
In this work, we use electric scooters (e-scooters) as an example to study the problem of how to incorporate city regulations (e.g., equity requirements and demand satisfaction) into shared micromobility vehicle scheduling.

The literature on multi-fleet management generally falls into two categories: (i) cooperative management~\cite{tan2023joint, pan2019deep, zhang2022multi, luo2021rebalancing} and (ii) non-cooperative management~\cite{wang2021record, lin2018efficient, li2021dynamic, yuan2019p}.
In the cooperative paradigm, some studies assume that all fleets share a common objective and work collectively to achieve it~\cite{tan2023joint}. However, this assumption is often unrealistic in practice, as micromobility operators pursue different goals driven by their own profit incentives~\cite{aarhaug2023price}, limiting the applicability of such cooperative frameworks.
Other cooperative approaches rely on a centralized controller~\cite{kondor2022cost} to coordinate vehicles from multiple operators. This setup faces practical barriers: operators may be unwilling to comply with centralized directives and typically prefer to retain control over their fleets. Moreover, a centralized control agent introduces a large action and state space, making the problem computationally challenging.
In contrast, non-cooperative frameworks model each fleet as operating independently with its own objectives, without centralized coordination~\cite{wang2021record}. While this setup reflects current practices more accurately, it precludes communication and cooperation between operators, making it difficult to achieve system-level outcomes that align with policy goals.
Although some real-world interventions~\cite{chicago_rule, escooter_guide} enforce regulations through strict rules (e.g., requiring a minimum number of vehicles in specific areas), these measures still fall short of meeting city-wide goals, as shown in Sec.~\ref{sec:Motivation}.

In our work, we design a novel approach to align the objectives of micromobility operators with broader city goals through a non-intrusive, collaborative mechanism.
We position the city regulator as a facilitator who interacts with these operators to identify optimal scheduling strategies that are aligned with city goals. This process involves operators continuously optimizing their fleet scheduling to maximize profitability, while the city regulator provides feedback that nudges their strategies toward city-wide goals. This feedback loop encourages operators to adjust their decisions without requiring direct control or intervention. This iterative process of adaptation and feedback facilitates a dynamic refinement of both the operators' scheduling actions and the city regulatory approaches.
Importantly, our mechanism allows the regulator to guide operator behavior solely through feedback, without access to the internal details (e.g., network structure or parameters) of their scheduling models. This preserves business confidentiality and supports practical deployment.
The goal is to establish a synergistic environment where micromobility services are optimized for both operator profitability and public objectives, such as promoting geographic equity and satisfying region-level demand~\cite{chicago_rule}.

We face two challenges in achieving our goal. 
Firstly, although micromobility operators are expected to comply with city regulations (such as those enforced in cities like Chicago, where violations may lead to permit suspension~\cite{chicago_rule}), the non-cooperative relationship between operators makes them unwilling to share real-time geographic vehicle locations and trip records with each other due to privacy and business competition.
As a result, each operator designs its scheduling policy based solely on its own observations.
Secondly, in current practice, operators are typically required to share only limited information, such as vehicle locations, with city regulators~\cite{escooter_guide}, without disclosing detailed model parameters or implementation details. This makes it difficult for the regulator to generate effective feedback.
Moreover, designing a fair feedback mechanism is itself a challenge. Since the feedback may affect each operator's ability to maximize profit, it is critical to ensure that the mechanism does not disproportionately advantage or disadvantage any individual operator, as our evaluation later demonstrates.

We design REALISM, a \underline{re}gul\underline{a}tory framework for coordinated schedu\underline{li}ng in multi-operator \underline{s}hared \underline{m}icromobility services. 
In our framework, we assume that each operator uses a scheduler built on a state-of-the-art algorithm, such as a Multi-Agent Reinforcement Learning (MARL) vehicle scheduling framework~\cite{tan2023joint, tan2024robust}, trained on its own data to maximize profit.
To guide these schedulers toward city-wide goals, we introduce a regulation network that evaluates the joint actions of all operators with respect to city regulations. 
It assigns individual rewards (positive or negative) to each operator based on their relative contributions to the city’s objectives.
To comprehensively measure the operators' region-level contributions, we introduce the Shapley value~\cite{shapley1953value, hu2022shape}, which provides a principled way to attribute the collective achievement of city goals to individual operators.
To ensure fairness in feedback, we incorporate multiple fairness metrics into the regulation network as a core component.
We further develop an adaptive alternating optimization procedure, allowing operator schedulers and the regulation model to iteratively interact and co-adapt until convergence.
The key contributions of this work are as follows:

(1) To our best knowledge, we are the first to approach the problem of multi-operator vehicle scheduling with city non-intrusive regulations.
It enables no shared information between operators and uses a feedback mechanism for regulators to guide the individual operators' scheduling.

(2) Technically, we design a regulatory framework for coordinated scheduling in multi-operator shared micromobility services. Specifically, the regulator scores the operators' rebalancing strategies and offers different rewards to them according to the city goal achievement and the operators' contribution, which is measured by the Shapley value of each operator. In addition, we develop an alternating procedure to make operators and the city regulator interact with each other to optimize the whole framework.

(3) We evaluate our framework based on real-world e-scooter usage data in Chicago. Our experiment results show that our method achieves a performance gain of at least 39.93$\%$ in the equity of vehicle usage and 1.82$\%$ in the average demand satisfaction of the whole city, compared to state-of-the-art baselines. On average, our method improves the equity of vehicle usage by 81.11$\%$ and the average demand satisfaction by 9.84$\%$ compared to those baselines.


\section{Preliminary and Motivation}
\label{sec:Motivation}
In this section, we first introduce the shared micromobility operation data. 
Then we motivate our work by analyzing the importance of city regulations on vehicle scheduling.

\begin{table}[h]
\caption{\small Samples in the dataset}
\vspace{-5pt}
\label{tb:data}
\resizebox{\linewidth}{!}{
\begin{tabular}{cccc}
\toprule
Trip ID          & Start Time       & End Time   & Trip Distance (m)                                                                                                                                  \\ 
T001     & 5/28/2022 14:00  & 5/28/2022 15:00  & 2,484  \\ \midrule
Trip Duration (s)   &  Start Region     & End Region       & Vehicle Operator                                                                                                                         \\ 
1,544 & -87.62519, 41.87887 & -87.62520 41.87886 & Lime  \\
\bottomrule 
\end{tabular}
}
\end{table}
\vspace{-5pt}

\subsection{Data Description}
\label{subsec:data}
We use a public real-world dataset collected by the Chicago government \cite{chicago_data} from 3 operators (Lime, Spin, and Bird) spanning four months from June 2022 to September 2022 with over 629,934 trips.
The dataset includes the trip duration, trip distance, vehicle operator, start time, start region, and other relevant information (as shown in Table~\ref{tb:data}).
The Chicago regulator (e.g., the transportation department) periodically gathers this information from operators.

\subsection{Why City Regulations Are Important?}
\label{sec:regulations}

\textbf{City Regulations}: One objective of the city regulator is to achieve the city goals through regulations.
For example, in Chicago, the city regulator requires shared micromobility operators to ensure vehicle usage equity by maintaining a minimum number of vehicles in high-priority regions \cite{chicago_rule} and maximizing the demand satisfaction rates of all areas~\cite{escooter_guide}.
In our work, the demand satisfaction rate is defined as the average ratio of satisfied demand to total demand across all regions per day. 
We assume that the number of trips recorded in the dataset is the total demand (this is commonly used as a reasonable approximation in existing works \cite{wang2021record,li2021dynamic,tan2023joint} when the actual demand is generally unknown).
The vehicle usage equity is the difference in supply-demand ratios between individual regions and the whole city (detailed formulation in Section~\ref{sec:problem}). 
A common city regulation is to define areas of a city with high priority, considering both demand satisfaction and vehicle usage equity, and instruct operators to relocate a fixed proportion of vehicles to those areas~\cite{chicago_rule}.

\begin{figure}[h]\centering
\begin{minipage}[h]{0.45\linewidth}
    \includegraphics[width=\linewidth, keepaspectratio=true]{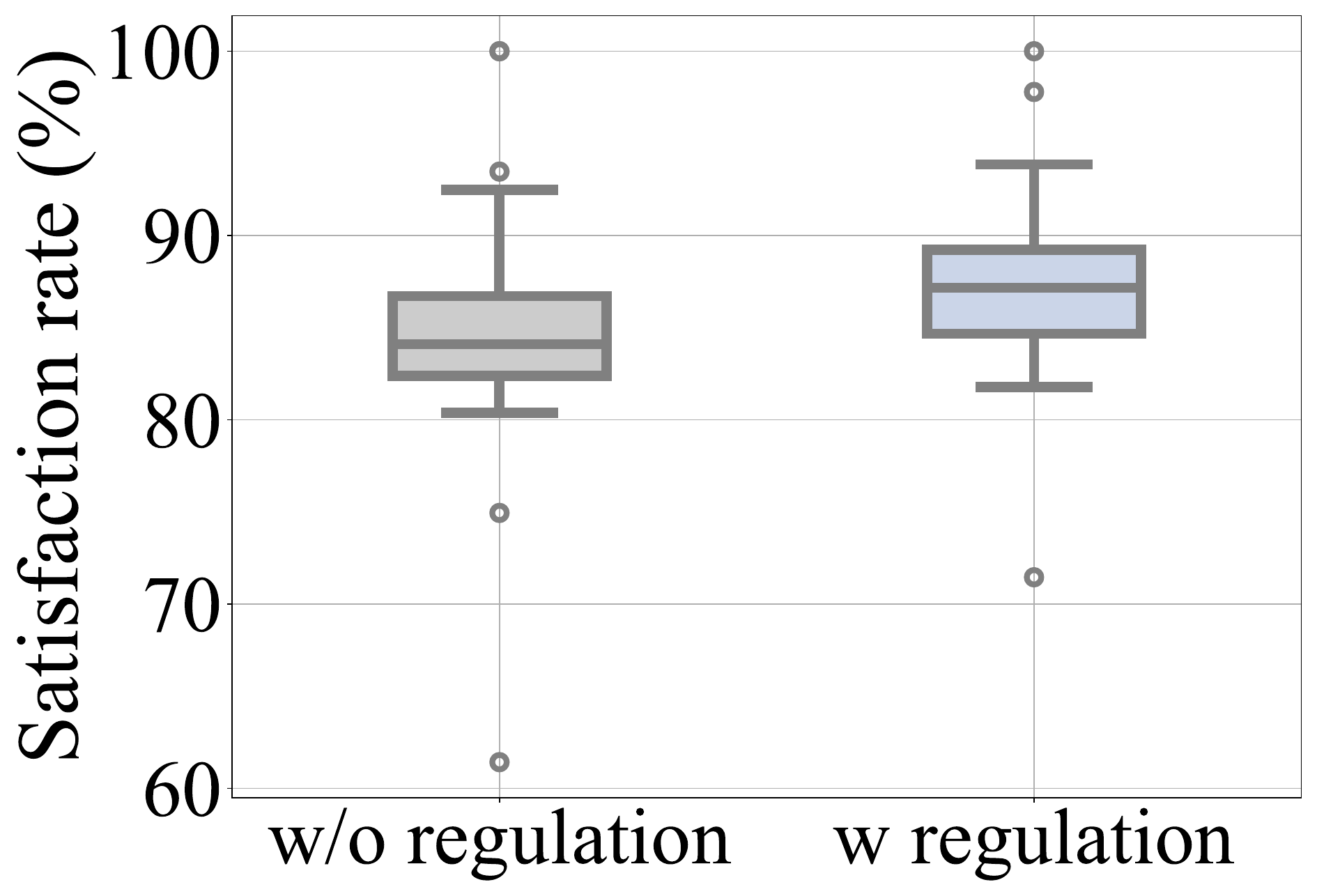}
    \vspace{-15pt}
    \captionsetup{font={small}}
    \caption{Distribution of daily demand satisfaction rate}
    \label{fig:motivation_sat}
\end{minipage}
\hspace{10pt}
\begin{minipage}[h]{0.45\linewidth}
    \includegraphics[width=\linewidth, keepaspectratio=true]{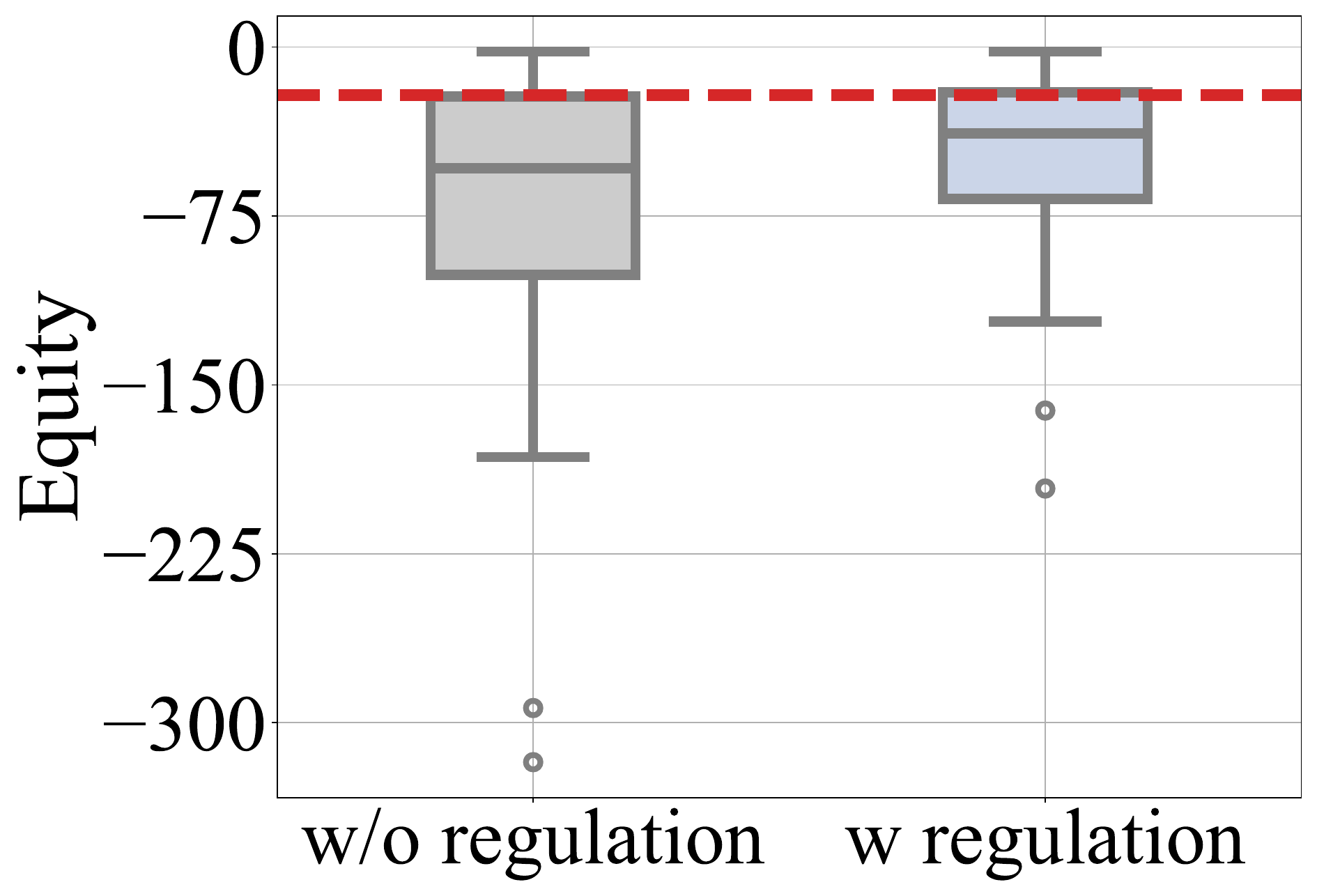}
    \vspace{-15pt}
    \captionsetup{font={small}}
    \caption{Distribution of daily vehicle usage equity}
    \label{fig:motivation_equ}
\end{minipage}
\vspace{-8pt}
\end{figure}

\textbf{Observations}: Our research is built upon the hypothesis that city regulations of vehicle scheduling are effective in fulfilling city goals.s
We examine the hypothesis by evaluating the achievement of the above city goals when applying scheduling methods with and without the implementation of city regulations based on the data in Chicago.
Based on the Chicago regulations~\cite{tan2023joint}, we relocate half of each operator's fleet to priority regions. 
Then, operators rebalance their remaining fleets based on their policies.
Figures~\ref{fig:motivation_sat} and~\ref{fig:motivation_equ} show the distributions of daily demand satisfaction rate and vehicle usage equity, jointly across three operators, in all 77 regions of the city over one month.
The red dashed line indicates the quantified city goal in Chicago~\cite{chicago_rule}.

As shown in Figure~\ref {fig:motivation_sat}, with regulation, the whole city can achieve a higher average demand satisfaction rate than that without regulation.
Then, Figure~\ref {fig:motivation_equ} shows that vehicle usage equity significantly falls below the city goal without regulation.
However, even if the regulation is applied to each operator, the improvement is marginal.
In addition, we find that the current city regulations (simply adding restrictions to each operator) undermine the operators' profits; for example, the net revenue of the operators Lime, Spin, and Bird decreased by 2.6$\%$, 3.3$\%$, and 4.4$\%$ respectively in our study. This implies that the operators' goal of maximizing their profits does not coincide with the city goal of maximizing the demand satisfaction rate due to the profit differences in trips.
Therefore, the above results motivate us to develop a more effective method to regulate all operators to satisfy city goals while optimizing operators' profits.

\section{Problem Formulation}
\label{sec:problem}
In this section, we formulate the problem of shared micromobility vehicle scheduling under regulations.

\textbf{Problem Setting}: We partition a city into $N$ regions according to the community divisions of the city~\cite{chicago_data}. A day is divided into $T$ equal-length time intervals. In the city, there are $M$ operators that provide shared micromobility services. To describe the spatial-temporal distribution of vehicles in a city, we use $S^{i,m}_{t}$ to denote the number of vehicles of operator $m$ in region $i$ at the beginning of time slot $t$ for $1\leq i\leq N,1\leq m\leq M$.
We use $U^{i,j,m}_{t}$ to denote user demand, which quantifies the number of user requests of operator $m$ from region $i$ to region $j$ within the time slot $t$ .
Therefore, for each operator, we can define the vehicle distribution and user requests of the whole city at time slot $t$ as $S^{m}_{t} \in \mathbb{N}^{N}$ $(S^{m}_{t}=\{S^{i,m}_{t}\}_{i,m})$, and $U^{m}_{t} \in \mathbb{N}^{N \times N}$ $(U^{m}_{t}=\{U^{i,j,m}_{t}\}_{i,j,m})$.

\textbf{Scheduling}: In order to meet the highest possible number of future user requests, it is essential for operators to rebalance vehicles.
 We define $a^{m}_{t} \in \mathbb{N}^{N \times N}$ $(a^{m}_{t} = \{a^{i,j,m}_{t}\}_{1\leq i,j\leq N})$ as the rebalancing strategy of operator $m$ at the beginning of time slot $t$, and $a^{i,j,m}_{t}$ represents the number of vehicles of operator $m$ to be rebalanced from region $i$ to $j$ at the beginning of time slot $t$.
To efficiently rebalance vehicles, operators should consider current vehicle distribution and user demand of future $h$ time slots. Hence, the rebalancing strategy $a^{m}_{t}$ is formulated as:
\begin{equation}
    a^{m}_{t} = f_{a,reb}(S^{m}_{t}, U^{m}_{t:t+h}).
\end{equation}
After determining rebalancing actions, the staff of each operator follow those actions and rebalance the vehicles by trucks. We define rebalancing costs $r^{reb,m}_{t}$ of each operator $m$ at time slot $t$ as:
\begin{equation}
    r^{reb,m}_{t} = f_{r,reb}(S^{m}_{t}, a^{m}_{t}).
\end{equation}
We use $Z^{m}_{t}$ to represent the trip revenue at time slot $t$, which is related to the vehicle distribution, user demand, and rebalancing actions of operator $m$ at time slot $t$ and formulate it as:
\begin{equation}
    Z^{m}_{t} = f_{trip}(S^{m}_{t}, U^{m}_{t}, a^{m}_{t}).
\end{equation}
\par We list the calculation details of trip revenue and truck-based rebalancing costs in the evaluation section (see Section~\ref{sec:evaluation}).

\textbf{City Goals}: The city regulator aims to realize specific social objectives by overseeing the regulation of shared micromobility vehicles throughout the city. In our study, we use Chicago as an example and identify two primary goals from the city regulations~\cite{city_goal}, which include (1) ensuring equitable vehicle usage across different communities, and (2) maximizing user demand satisfaction in all areas. It is worth noting that other goals can also be incorporated into our framework with minimal modifications to our method.
Motivated by~\cite{he2023robust,miao2015taxi}, we define the equity of vehicle usage across the regions of a city as the difference between the region-level demand-supply ratio (${U^{i}_{t}}/{S^{i}_{t}}$) and the city-level demand-supply ratio (i.e., ${\sum^{N}_{i'=1}U^{i'}_{t}}/{\sum^{N}_{i'=1}S^{i'}_{t}}$), formulated as:
\begin{equation}
\label{eq:equ}
    C^{equ}_{t} = \sum^{N}_{i=1} - \Big|  \frac{U^{i}_{t}}{S^{i}_{t}} - \frac{\sum^{N}_{i'=1}U^{i'}_{t}}{\sum^{N}_{i'=1}S^{i'}_{t}} \Big|,
\end{equation}
where $S^{i}_{t}$ is the number of vehicles in region $i$ at time slot $t$. 
We define $D_{t} = \{D^{i}_{t}\}_i$ as the number of satisfied demands of the region $i$ at time slot $t$, and $U_{t} = \{U^{i}_{t}\}_i$ as the number of user demand in the region at the beginning of time slot $t$. 
The satisfaction rate of user demand in the city $C^{sat}_{t}$ is formulated as:
\begin{equation}
\label{eq:sat}
    C^{sat}_{t} = \frac{1}{N} \cdot \sum^{N}_{i=1}\frac{D^{i}_{t}}{U^{i}_{t}}.
\end{equation}
Based on the above two metrics, the city goals are defined as:
\begin{equation}
\label{eq:constraint}
    C^{equ}_{t} > Q_{equ}, \quad C^{sat}_{t} > Q_{sat}, \quad t=1,\cdots,T,
\end{equation}
where $Q_{equ}$ and $Q_{sat}$ are the constant requirements of satisfaction rate and vehicle usage equity for the entire city at each time slot.

\textbf{Objective}: 
The operators and the city regulator have inconsistent objectives of shared micromobility services.
From the operator's perspective, its goal is to develop algorithms for efficiently determining the scheduling of vehicles to maximize the net revenue $R_{total}$ (i.e., the total trip revenue minus the rebalancing costs). 
We define the objective function of a micromobility operator as:
\begin{align}
        \max_{a^{m}_{t}} &  \ \ \ R_{total} = \sum^{T}_{t=1}f_{trip}(S^{m}_{t}, U^{m}_{t}, a^{m}_{t}) - \sum^{T}_{t=1}f_{r,reb}(S^{m}_{t}, a^{m}_{t}) \\
   & \textbf{s.t.}\ \  \; C^{sat}_{t} > Q_{sat}, \quad C^{equ}_{t} > Q_{equ}, \quad t=1,\cdots,T. \nonumber
\end{align}
The city regulator seeks to foster collaboration among operators to attain the city goals, which are formulated as the constraints in the above equation.


\section{Methodology}

\begin{figure}[h]
    \centering
    \includegraphics[width=1.0\linewidth, keepaspectratio=true]{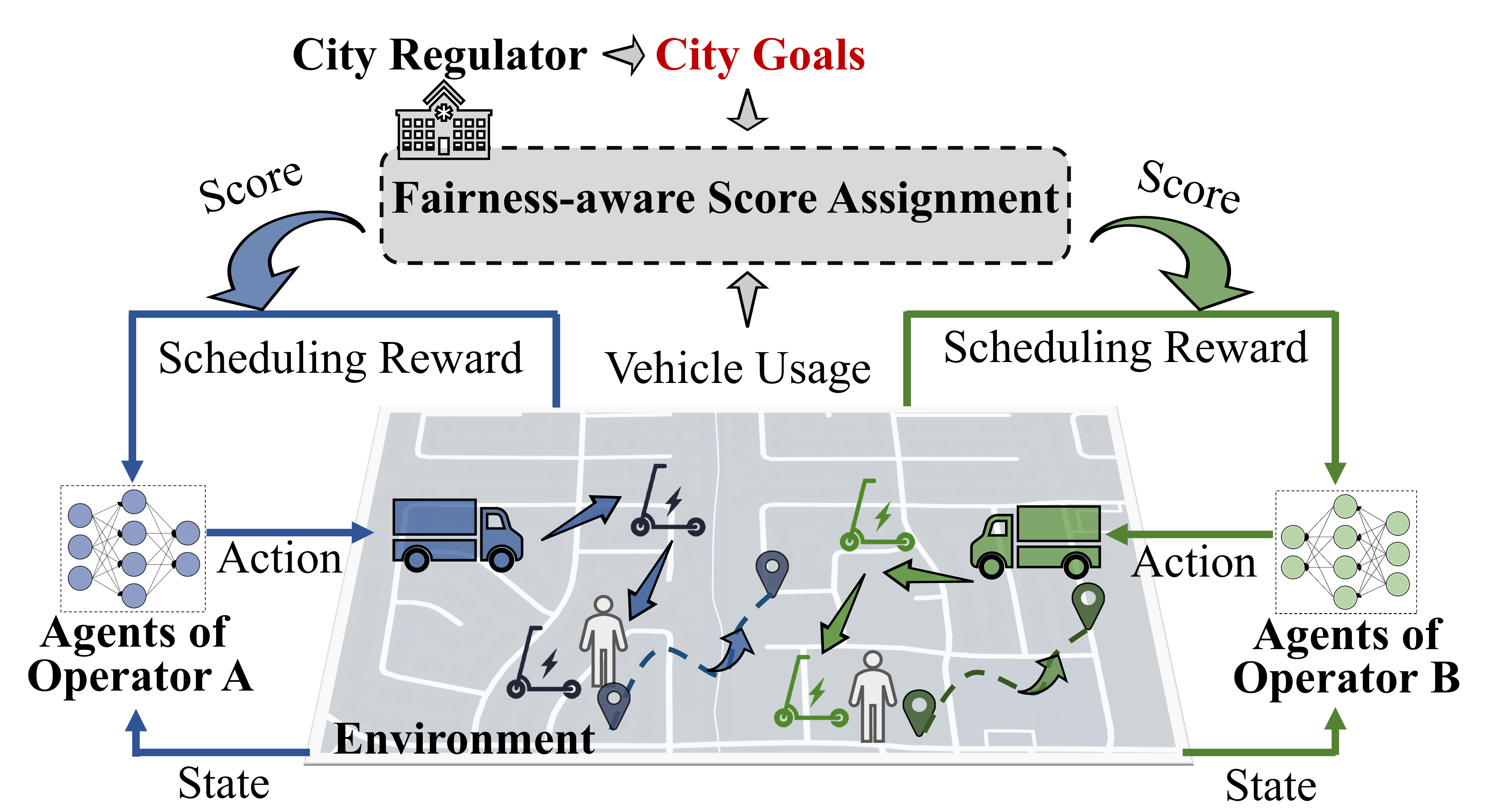}
    \vspace{-15pt}
    \caption{Overview of MARL framework for rebalancing shared micromobility vehicles under city regulations}
    \label{fig:framework}
    \vspace{-15pt}
\end{figure}

\subsection{Design Overview}
\label{subsec:overview}
We design a MARL-based framework for rebalancing shared micromobility vehicles under the city regulations, as shown in Fig~\ref{fig:framework}. This framework consists of three components: the environment, shared micromobility operators, and a city regulator. (1) \textbf{Environment:} It simulates the shared micromobility system operations. (2) \textbf{Shared micromobility operators:} In our work, we assume that users rely on a single operator—they do not switch between operators when their preferred operator has insufficient supply. Each operator is assumed to utilize a neural network-based scheduling model and learn how to schedule its own vehicles in the city to maximize the net revenue, determined by the balance between satisfied user demand and rebalancing costs. Specifically, we assign a neural network-based agent for each operator in a region to make rebalancing decisions considering the vehicle distribution and future user demand. 
The rebalancing strategies are then executed by trucks that move vehicles between communities. 
(3) \textbf{City regulator:} In practice, cities like Chicago regulate shared micromobility by requiring operators to share a real-time data API~\cite{chicago_rule}. Incentives and penalties—such as (i) suspending operations for up to 30 days and (ii) adjusting fleet sizes based on performance—are imposed based on the regulator-defined shared micromobility service metrics, such as~\cite{chicago_rule}, which have direct financial implications for operators. Therefore, in the city regulator's component, we consider the city regulator as a trusted entity to which operators share real-time mobility data and assess the progress toward meeting the city goals. Then, the city regulator offers a financial score, which can be either negative (penalty) or positive (incentive), to each operator, taking into account their respective contributions to the fulfillment of the city goals.
In our study, we design a fairness-aware score assignment model that processes inputs such as the city goal achievements, each operator's contribution towards these goals, and the daily net revenue of the operators.
The model outputs a score and assigns it to the reward for each operator to update its rebalancing agents.
Through the implementation of fairness-aware score assignments, we expect operators to collaboratively fulfill the city goals while also optimizing their profits.

\subsection{Multi-agent RL-based Vehicle Scheduling}
\label{MARL}
Motivated by the advancement of vehicle scheduling methods~\cite{tan2023joint,wang2021record}, we model the problem of rebalancing vehicles for each operator as a Markov Decision Process for $N$ agents expressed as a tuple ($\{\mathcal{S}, \mathcal{A}, \mathcal{R}, \mathcal{P}, \gamma\}$). $\mathcal{S}$ denotes the set of states. $\mathcal{A}$ represents the set of action. $\mathcal{P}: \mathcal{S} \times \mathcal{A} \times \mathcal{S} \rightarrow [0,1]$ denotes the transition probability. $\mathcal{R}$ is the reward function. $\gamma$ denotes the discounted factor. The definitions of these notations are as follows:

\textbf{Agent}: We define an agent for each region in the city, which decides how to rebalance the vehicles in a region. Instead of utilizing a single agent to control the vehicles for the entire city, we use centralized training and decentralized execution to reduce the computational complexity~\cite{lowe2017multi, he2023robust}.

\textbf{State}: At the beginning of each time slot $t$, the state of region $i$ is defined as $s^{i}_{t} = \{s^{lo, i}_{t}, s^{gl}_{t}\}$, where the local state $s^{lo,i}_{t}$ contains the number of vehicles $S^{i}_{t}$ and future $h$ time-slot user demand $U^{i}_{t:t+h}$ $(U^{i}_{t:t+h} = \{U^{i,j}_{t:t+h}\}_j)$ in region $i$ at the beginning of time slot $t$. The global state $s^{gl}_{t}$ contains the vehicle distribution $S_{t}$ and the city-level user demand during the future $h$ time slots $U_{t:t+h}$. The future demand is predicted by a pre-trained prediction model~\cite{tan2023joint}.

\textbf{Action}: Given the above state, the agent in region $i$ at the beginning of time slot $t$ decides the number of vehicles to be rebalanced from region $i$ to  $j$, denoted as $a^{i}_{t} = \{a^{i,j}_{t}\}_{1\leq j\leq N})$.

\textbf{Reward}: Since all the region agents collaboratively maximize the operator profit, they share the same reward. In our work, the reward is defined as the net revenue, consisting of trip revenue from serving users, rebalancing costs for truck movements, and the score assigned by the city regulator, formulated as:
\begin{equation}
\label{eq:reward}
    r_{t} = Z_{t} - \alpha{} \cdot Z^{reb}_{t} - Z^{pen}_{t},
\end{equation}
where $Z_{t}$ denotes the revenue generated from trips, while $Z^{reb}_{t}$ represents the cost associated with truck mileage during time slot $t$, determined through a truck-routing optimization algorithm \cite{tan2023joint, li2021dynamic}. $\alpha$ serves as a conversion coefficient, translating various values into monetary terms. $Z^{pen}_{t}$ is the financial score imposed by the city regulator at the conclusion of time slot $t$, with further details provided in Section~\ref{fairness}.

\textbf{Transition probability function}: It denotes the probability that the joint state $s_{t}$ $(s_{t}=\{s^{i}_{t}\}_i)$ transfers to the next joint state $s_{t+1}$ given the joint action $a_t$ $(a_{t}=\{a^{i}_{t}\}_i)$.

\textbf{Discounted Factor}: The discounted factor $\gamma$ signifies the degree to which the agents prioritize the future reward over immediate ones, $\gamma \in [0,1)$. A $\gamma$ value of 0 indicates that the agent prioritizes immediate rewards exclusively, thus directing its learning towards actions that yield instant benefits.

Given the above settings, the objective of all the region agents for each operator is to collaboratively maximize the expected cumulative reward, denoted as $G_{t} = [\sum^{\infty}_{t=1}\gamma^{t-1}R(s_{t}, a_{t}|s_{1}=s)]$. The Q-value of joint state $s_{t}$ and action $a_{t}$ under policy $\pi_{\theta}$ is denoted by $Q^{\pi_{\theta}}=E[\sum^{\infty}_{k=0}$ $R(s_{t+k+1}, a_{t+k+1})|\pi_{\theta, s_{t}, a_{t}}]$.

\subsection{Fairness-aware Score Assignment}
\label{fairness}

City regulators primarily focus on the fulfillment of city goals. The contributions of various operators towards these goals vary, influenced by inherent operator characteristics (e.g., fleet sizes) and operational strategies. 
In our work, city regulators take into account three critical factors when determining score assignments: the extent of city goal achievements, the varied contributions of different operators towards these goals, and the respective net revenues of these operators.

\textbf{City goal achievement}: We evaluate the city goal achievement by comparing the city's micromobility status at each time slot with the specified requirements of city goals.
This measurement is based on Eq. \eqref{eq:constraint} and is formulated as follows:

\begin{equation}
\begin{aligned}
    G^{sat}_{t} &= Q_{sat} - \frac{1}{N} \cdot \sum^{N}_{i=1}\frac{D^{i}_{t}}{U^{i}_{t}};\\
    G^{equ}_{t} &= Q_{equ} - \sum^{N}_{i=1} - \Big| \frac{U^{i}_{t}}{S^{i}_{t}} - \frac{\sum^{N}_{i'=1}U^{i'}_{t}}{\sum^{N}_{i'=1}S^{i'}_{t}}  \Big|,
\end{aligned}
\end{equation}
where $G^{sat}_{t}$ and $G^{equ}_{t}$ represent the distance between the achieved demand satisfaction and vehicle usage equity at time slot $t$, and the city goals (i.e., $Q_{sat}$ and $Q_{equ}$), respectively. 

\textbf{Operator contribution}: 
Due to variations in vehicle distribution and user demand across different operators, merely assessing an operator's contribution to city goals through Eq. \eqref{eq:sat} for satisfaction and Eq. \eqref{eq:equ} for equity does not accurately capture the operator's true impact. 
Inspired by foundational works in coalitional game theory~\cite{hu2022shape, shapley1953value}, we employ the Shapley value as a metric to quantify each operator's individual contribution.
We conceptualize the calculation of an operator's contribution as a game involving multiple operators. We denote the set of operators participating in this game as $\mathbf{M}:=\{1,..., M\}$. Within this framework, operators have the potential to form various coalitions, exemplified by subsets of $\mathbf{M}$ such as $\{1, 2\}$.
Considering an operator $m$ and a subset $H \subseteq \mathbf{M}\symbol{92}\{m\}$, the marginal contribution of $m$ to $H$ from the perspectives of demand satisfaction and vehicle usage equity at time slot $t$ are evaluated by:
\begin{equation}
    \hat{C}^{sat}_{m,t}(H;C^{sat}_{t}) = C^{sat}_{t}(H\cup \{m\}) - C^{sat}_{t}(H),
\end{equation}
\begin{equation}
    \hat{C}^{equ}_{m,t}(H;C^{equ}_{t}) = C^{equ}_{t}(H\cup \{m\}) - C^{equ}_{t}(H),
\end{equation}
where $C^{sat}_{t}(H)$ and $C^{equ}_{t}(H)$ are the city goal achievements considering the vehicle distribution and user demand of operator subset $H$ at time slot $t$. 
The marginal contribution represents the contribution of the operator $m$ to the operator subset $H$ from the perspective of one city goal achievement.
Since an operator can interact with any other operators, the marginal contribution of the operator $i$ should consider all possible $\hat{C}^{equ}_{m,t}$ and $\hat{C}^{sat}_{m,t}$. Therefore, the Shaley value of operator $m$ from the perspectives of demand satisfaction and usage equity at time slot $t$ is defined as:
\begin{equation}
    \phi^{sat,m}_{t}(M;C^{sat}_{t}) = \sum_{H \subseteq \mathbf{M}\symbol{92}\{m\}} \frac{|H|!(M-|H|-1)!}{M!}\hat{C}^{sat}_{m,t},
\end{equation}
\begin{equation}
    \phi^{equ,m}_{t}(M;C^{equ}_{t}) = \sum_{H \subseteq \mathbf{M}\symbol{92}\{m\}} \frac{|H|!(M-|H|-1)!}{M!}\hat{C}^{equ}_{m,t}.
\end{equation}
By calculating the Shapley value of operator $i$ from the perspectives of demand satisfaction and usage equity, the city regulator can know the operator's individual contribution to each city goal.

\renewcommand{\algorithmicrequire}{\textbf{Input:}}
\renewcommand{\algorithmicensure}{\textbf{Initialize:}}

\begin{algorithm}
\small
    \caption{MARL for rebalancing shared micromobility vehicles under city regulator's regulations}
    \label{ab1}
    \begin{algorithmic}
    \REQUIRE Environment; $\alpha$; $\tau$; $\beta$, noise clip $c$
    \ENSURE Region agents actor and critic policies of different operators $\theta^{m,i}$, $\psi^{m,i}$, score assignment prediction network $\delta$
    \FOR{$n = 1$ to $N_{iter}$}
    \STATE Receive initial state s
    \STATE \textbf{/* Task 1: Collect and store trajectories in the buffer based on the previous strategy. */}
    \FOR{$t=1$ to episode-length $T$}
    \STATE Each operator agent selects an action based on its observation $o^{m}_{i}$ with exploration noise $\epsilon$: $a^{m,i}_{t} = \pi_{\theta^{m,i}}(o^{m}_{i})+\epsilon$, $\epsilon \sim \mathcal{N}(0, \sigma)$
    \STATE Execute operators' actions $a^{m} = \{a^{m}_{i}\}$, $\forall m \in M$, $\forall i \in N$
    \STATE The city regulator assigns a score to each operator by observing the shared micromobility in the city $P_{t} = f_{pen}(G^{sat}_{t}, G^{equ}_{t}, \phi^{sat}_{t}, \phi^{equ}_{t}, Z^{net}_{t})$, $r_{t} = Z_{t} - \alpha{} \cdot Z^{reb}_{t} - Z^{pen}_{t}$.
    \STATE Calculate the fairness of score assignment:
    \STATE $E_{t} = \sum^{M}_{i=1} -|\frac{Z^{pen,m}_{t}}{Z^{net,m}_{t}}-\frac{\sum^{M}_{i=1}Z^{pen,m}_{t}}{\sum^{M}_{i=1}Z^{net,m}_{t}}|$
    \STATE Store $(s, a_{t}, r_{t}, s')$ in replay buffer $\mathcal{D}$
    \STATE $s$ $\leftarrow$ $s'$
    \ENDFOR
    \STATE \textbf{/* Task 2: Update the region agents of all operators by sampling trajectories from the buffer. */}
    \FOR{operator $m=1$ to M}
    \FOR{agent $i=1$ to N}
    \STATE Sample a random minibatch of $\mathcal{X}$ samples $(s^{m,j}, a^{m,j}, r^{m,j}, s'^{m,j})$ from $\mathcal{D}$
    \STATE set $y^{j}=r^{j}_{m,i}+$
    \STATE $\gamma Q^{\psi'}_{m,i}(s'^{m,j}, \{a'^{m}_{i}\})|_{a'^{m}_{i}=\pi_{\theta^{m,i}}(o'^{m,j}_{i})+\epsilon}$, 
    \STATE $\epsilon \sim clip(\mathcal{N}(0, \sigma), -c, c)$
    \STATE Update agent critic policy
    \STATE $\psi^{m,i} \leftarrow \arg\min_{\psi^{m,i}} \frac{1}{\mathcal{X}} \sum_{j}(y^{j}$
    \STATE $\quad -Q^{\psi}_{m,i}(s, \{a^{m}_{i}\}))^{2}$
    \STATE Update agent actor policy
    \STATE $\nabla_{\theta^{m,i}}\mathcal{J} \approx \frac{1}{\mathcal{X}}\sum_{j}\nabla_{\theta^{m,i}}\pi_{\theta^{m,i}}(o^{m,j}_{i})$
    \STATE $\quad \nabla_{a^{m}_{i}}Q^{\psi}_{m,i}(s^{j}, \{a^{m}_{i}\})|_{a^{m}_{i}}$
    \ENDFOR
    \ENDFOR
    \STATE Update target networks
    \STATE $\theta'^{m,i} \leftarrow \tau \theta^{m,i} + (1-\tau)\theta'^{m,i}$
    \STATE $\psi'^{m,i} \leftarrow \tau \psi^{m,i} + (1-\tau)\psi'^{m,i}$
    \STATE \textbf{/* Task 3: Update the fairness-aware score assignment modes by replaying the iteration from task 2 with the same initial state but using the updated rebalancing policies. */}
    \STATE Initialize the state $s$, and replay the iteration.
    \FOR{$t'=1$ to episode-length $T$}
    \STATE Collecting the achievement distance of city goals ($G^{sat}_{t'}$, $G^{equ}_{t'}$) for each time slot $t'$
    \ENDFOR
    \STATE Calculate the loss of score assignment prediction based on achievement distance and average fairness of score assignment in the previous round:
    \STATE $\mathcal{L}_{\delta}=\beta \frac{1}{T}\sum_{t'}(G^{sat}_{t'} + G^{equ}_{t'}) + (1-\beta)\frac{1}{T}\sum_{t}E_{t}$
    \STATE Update the prediction network $\delta$
    \ENDFOR
    \RETURN $\{\theta^{m,i}\}$, $\{\psi^{m,i}\}$, $\delta$
    \end{algorithmic}
\end{algorithm}

\textbf{Score assignment}: After identifying the city goal achievement ($G^{sat}_{t}$ and $G^{equ}_{t}$), and the operator's individual contribution $(\phi^{sat,m}_{t}$ and $\phi^{equ,m}_{t}$, $m=1,...,M)$, we develop a neural network-based model to assign a score to each operator 
$(Z^{pen}_{t}=\{Z^{pen,m}_{t}\}_m$. 
The generated scores of the developed model is represented as:
\begin{equation}
    Z^{pen}_{t} = f_{pen}(G^{sat}_{t}, G^{equ}_{t}, \phi^{sat}_{t}, \phi^{equ}_{t}, Z^{net}_{t}),
\end{equation}
where $Z^{net}_{t} = \{Z^{net,m}_{t}\}_m$ is the set of operators' net revenue at the time slot $t$. The operator's net revenue is calculated based on the trip revenue from satisfied demand and rebalancing costs (detailed description in Subsection \ref{MARL}):
\begin{equation}
    Z^{net,m}_{t} = Z^{m}_{t} - \alpha \cdot Z^{reb, m}_{t}.
\end{equation}
After the score is assigned to each operator, the city operate feeds the score to its reward function (seen in Eq. \eqref{eq:reward}).

Considering the score is fed into the reward function and can impact each operator's profit, it is important for the city regulator to achieve fairness-aware score assignment. 
In our work, we evaluate the fairness of score assignment by measuring the difference of the score-revenue ratios between individual operators and all the operators, motivated by the existing work \cite{he2023robust}, formulated as:
\begin{equation}
\label{eq:fairness}
    E_{t} = \sum^{M}_{m=1} -|\frac{Z^{pen,m}_{t}}{Z^{net,m}_{t}}-\frac{\sum^{M}_{m'=1}Z^{pen,m'}_{t}}{\sum^{M}_{m'=1}Z^{net,m'}_{t}}|.
\end{equation}
As different definitions of fairness emphasize varied aspects, we also evaluate how these definitions affect performance in the evaluation section (see Section \ref{sec:evaluation}).

\subsection{Training Method}
To optimize the agents of different operators and the fairness-aware score assignment model, we adopt a two-step adaptive alternating procedure motivated by the existing Nash Equilibrium work \cite{pinto2017robust}.
We divide each iteration into three tasks, as shown in Algorithm~\ref{ab1}:
\noindent\textbf{Task 1: Trajectory collection:} 
We collect and store trajectories in the buffer based on the previous strategy. In this task, operators will interact with the regulator: agents of each operator rebalance vehicles according to their policies, and the city regulator assigns a score to each operator based on the city goal achievements and operators' contributions, affecting operators' reward functions. The operators' trajectories are stored in a data buffer.

\noindent\textbf{Task 2: Update operator agents:} 
Given a fixed fairness-aware score assignment model, we update the operators' agents through a standard reinforcement learning updating scheme.
During each training iteration, we collect actions, states, and rewards from the region agents to update their policies by existing gradient policy update methods~\cite{lowe2017multi, fujimoto2018addressing}.

\noindent \textbf{Task 3: Update the fairness-aware score assignment model:} 
After updating the operator agents, we proceed to update the score assignment model. 
In this task, we replay the interaction from task 2 with the same initial state but using the updated operators' policies, for further assessment.
As this model influences the rebalancing policy updates, which in turn affects both the city goal achievement and the operators' profits, we assess the effectiveness of the scores from two perspectives by replaying the iteration: (i) whether the rebalancing policy updates guided by the scores move the outcomes closer to the city goals, and (ii) whether the scores are fair to each operator.
Thus, we supervise the model update with these two objectives. 
For (i), we replay the iteration from step 1 with the same initial state but use the updated rebalancing policie to collect new data on city goal achievement, calculated as $\frac{1}{T}\sum_{t'}(G^{sat}_{t'} + G^{equ}_{t'})$, where T and $t'$ denote the episode length and each time slot, respectively. We use $t'$ to distinguish it from the time slot $t$ in step 1. 
An improvement in city goal achievement indicates that the score-guided rebalancing policy updates are more effectively aligned with the city goals. 
For (ii), we use Equation~\ref{eq:fairness} to measure the fairness of the scores assigned to operators.
Consequently, the loss function used to update the score assignment model is represented as:
\begin{equation}
    \mathcal{L}_{\delta}=\beta \frac{1}{T}\sum_{t'}(G^{sat}_{t'} + G^{equ}_{t'}) + (1-\beta)\frac{1}{T}\sum_{t}E_{t},
\end{equation}
where $\beta$ is a significance weight. 
Then, we continuously repeat these three tasks until convergence (e.g., achieving stable profits and city goal achievements).


\section{Evaluation}
\label{sec:evaluation}

\begin{table*}[t]
\caption{Performance comparison of different methods. The average improvements of REALISM over the four baselines across different metrics are shown in the bottom row.}
\label{tab:performance}
\vspace{-5pt}

\resizebox{\textwidth}{!}{
\begin{tabular}{c|ccc|cc}
\toprule
\multirow{2}{*}{Method} & \multicolumn{3}{c|}{Net Revenue (\$)}                                                                                                                                                                             & \multicolumn{2}{c}{City Goal Achievement}                                                \\ \cmidrule{2-6} 
                        & \multicolumn{1}{c}{Lime}  & \multicolumn{1}{c}{Spin} & Bird  & \multicolumn{1}{c}{Satisfaction Rate (\%)} & Vehicle Usage Equity \\ 
\midrule
SDSM    & 128,473.26 ($\pm$72.35)  & 263,866.39 ($\pm$40.74)  & 378,939.68 ($\pm$262.48)   & 68.24 ($\pm$0.44)   & -234.09                   \\ 
RECOMMEND \cite{tan2023joint}   & 241,789.25 ($\pm$1,207.78)  & 417,776.28 ($\pm$5,001.50)   & 480,096.86 ($\pm$5,420.85)  & 88.51 ($\pm$1.37)  & -56.03 ($\pm$1.35)      \\ 
ROCOMA \cite{he2023robust} & 238,415.89 ($\pm$1,237.91)  & 402,724.36 ($\pm$4,840.84)   & 455,357.58 ($\pm$5,463.40)  & 85.77 ($\pm$1.79)  & -27.78 ($\pm$1.84)  \\
MAC-A2C \cite{babaeizadeh2016reinforcement} & 224,657.02 ($\pm$1,108.81)  & 383,900.51 ($\pm$4,371.12)   & 434,126.75 ($\pm$5,142.21)  & 85.66 ($\pm$1.84)  & -27.12 ($\pm$1.62)  \\

\midrule
REALISM w/o R & 241,351.20 ($\pm$1,226.23)     & 417,788.72 ($\pm$5,094.36)    & 489,203.76 ($\pm$5,447.68) & 89.07 ($\pm$1.37)  & -49.74 ($\pm$1.35)            \\ 
REALISM w/o FASA & 248,660.70 ($\pm$1,352.19)     & 426,238.80 ($\pm$5,168.08)    & 502,660.68 ($\pm$5,755.58) & 89.49 ($\pm$1.74)  & -16.78 ($\pm$1.21)            \\\midrule 
Oracle & 258,209.78 ($\pm$1,682.67)  & 455,784.68 ($\pm$5,371.12)   & 543,536.28 ($\pm$5,999.96)  & 93.25 ($\pm$1.65)  & -12.53 ($\pm$1.59)  \\

\textbf{REALISM} & 243,107.87 ($\pm$1,243.41)     & 428,709.72 ($\pm$5,383.38)    & 507,418.75 ($\pm$5,812.73) & 90.12 ($\pm$1.36)  & -16.29 ($\pm$1.33)            \\ 
\midrule
Improvement & 16.69$\%$ & 16.79$\%$ & 16.08$\%$ & 9.84$\%$ & 81.11$\%$ \\
\bottomrule
\end{tabular}
}
\end{table*}

\subsection{Experimental Setup}

\subsubsection{\textbf{Implementation}} We conduct experiments on a public real-world shared e-scooter dataset in Chicago~\cite{chicago_data}. We divide the dataset into two parts: the data from the first two months serves as the training set, and the remaining data serves as the testing set. The entire city is partitioned into 77 regions based on the existing community divisions in the dataset. A day is divided into 24 hours, with operators' rebalancing occurring every 12 hours.
In our experiments, we consider three operators (Lime, Spin, and Bird), with the number of vehicles from each operator being 2,695, 2,581, and 2,795, respectively. 
The maximum reachable distance of an e-scooter is calculated based on its energy consumption and battery capacity~\cite{escooter_guide}. Only the nearby vehicles with enough remaining energy can satisfy the user demand, and the trip revenue is calculated at \$1.00 to unlock + \$0.39 per minute, excluding tax~\cite{chicago_price}. The truck traveling cost per kilometer is set at $\$2.422$, based on gas prices and truck fuel consumption from existing work~\cite{tan2023joint}. To simplify, we exclude the labor cost from our calculations. In the test stage, we use the updated operators' agents.

We implement our method and baselines with PyTorch 1.9.1, Python-mip 1.14.2, and gym 0.21.0 in a Python 3.7 environment, and we train them on a server equipped with 32 GB of memory and a GeForce RTX 3080 Ti GPU. By testing the performances under different hyperparameter settings, we use the following settings:
For operator agents, we use a stochastic gradient descent optimizer with a learning rate of 1e-4, determined to be optimal among [1e-3, 1e-4, 1e-5]. City regulation learning employs an Adam optimizer with an optimal learning rate of 1e-3 from the same range. Exploration noise is best at 0.1, tested among [0.05, 0.1, 0.2], with a minibatch size of 32 for experience replay. The discount factor $\gamma$ is set as 0.99. Hyperparameters for other baselines are fine-tuned based on the range in the original papers.
Each operator's multi-agent vehicle scheduling model is pre-trained based on the existing work \cite{tan2023joint}, and our regulatory framework for multi-operator vehicle scheduling converged in 124 iterations.

\subsubsection{\textbf{Baselines}} We compare our method with the following baselines and variants of our model:
\begin{itemize}
    \item \textbf{SDSM}. It is a static demand-supply matching method where each operator rebalances the vehicles to each region based on the ratio of the historical demand.
    \item \textbf{RECOMMEND} \cite{tan2023joint}. It is a state-of-the-art shared electric micromobility vehicle rebalancing and charging algorithm. We remove the components that involve charging while remaining the rebalancing components.
    \item \textbf{ROCOMA} \cite{he2023robust}. It is a state-of-the-art shared electric vehicle rebalancing algorithm to minimize both the rebalancing costs and the inequities of vehicle usage and charging aross all regions.
    \item \textbf{MAC-A2C} \cite{babaeizadeh2016reinforcement}. It is a fully cooperative multi-agent reinforcement learning framework adapted to collaboratively maximize the achievement of city goals.
    \item \textbf{Oracle}. It represents the upper-bound performance where the operators completely follow a trained centralized scheduler~\cite{tan2023joint} whose goal is to optimize the total revenue and the city goal achievement jointly. Designing Oracle is to show the gap between our method and the upper bound derived from an ideal (impractical) case.
\end{itemize}

Variants of our model are as follows:
\begin{itemize}
    \item \textbf{REALISM without Regulation (w/o R)}. In this setting, we remove the city regulator's regulation and let operators in the city maximize their own profit without considering the city goal achievement.
    \item \textbf{REALISM without Fairness-Awareness Score Assignment (w/o FASA)}. In this setting, we remove the fairness evaluation of the city regulator's score assignment. The loss function of the score assignment prediction model only considers the satisfaction rate.
    \item \textbf{REALISM with Different Fairness Definitions (w DFD)}. In this setting, we utilize a different definition of fairness~\cite{jiang2023faircod} from perspectives of both vehicle usage and score assignment. 
    Specifically, we use the difference between the local demand satisfaction rate and the global demand satisfaction rate, and the difference between individual operator score and average operator score to respectively measure the vehicle usage equity and the fairness of score assignment.
\end{itemize}

\subsubsection{\textbf{Metrics}} The evaluation metrics are as follows:
\begin{itemize}
    \item \textbf{Average satisfaction rate}: average satisfaction rate represents the average ratio of satisfied demand to the total user demand among all the regions (Equation \ref{eq:sat}).
    \item \textbf{Net revenue}: net revenue is the operator's trip revenue from satisfied user demand minus the rebalancing costs of the relevant operator (Equation \ref{eq:reward}).
    \item \textbf{Vehicle usage equity}: vehicle usage equity measures the difference between demand-supply ratios of individual regions and the whole city (Equation \ref{eq:equ}).
\end{itemize}

\subsection{Overall Performance}
Table~\ref{tab:performance} shows the overall performance of different methods, including the model Oracle, which reflects the upper-bound performance of our model. From this analysis, we have the following findings:
\textbf{(1) Overall, our model REALISM consistently outperforms other state-of-the-art methods}, achieving an improvement of at least 39.93\% in terms of vehicle usage equity and 1.82\% in satisfaction rate. 
\textbf{(2) Compared with the city goal-aware baselines (i.e., ROCOMA and MAC-A2C)}, REALISM significantly improves operators' net revenue. This indicates that the component of the city regulator's regulation can guide operators' rebalancing from a global perspective, enabling operators to provide more efficient and effective rebalancing strategies. Additionally, it is worth mentioning that the improvement in demand satisfaction rate is not as pronounced. This can be partially attributed to the profit differences among trips, suggesting that while operators strive to maximize net revenue and satisfaction rates, higher profits from individual trips do not necessarily translate to an increased demand satisfaction rate.
\textbf{(3) Compared with the baselines without considering city goal achievements (i.e., 
 SDSM and RECOMMEND)}, REALISM shows superiority in improving vehicle usage equity due to its awareness of city goal achievements.
\textbf{(4) Compared with the upper-bound model (i.e., Oracle)}, our model achieves comparably good performance. This demonstrates the advantage of our method under conditions where there is no information sharing between operators, making it practically useful. 
Even though Oracle achieves a higher performance, due to the competitive nature and data privacy concerns, it is generally considered an impractical setting in reality.

\subsection{Ablation Study}
\subsubsection{\textbf{The Effectiveness of Regulation}}
To demonstrate the effectiveness of the city regulator's regulation, we compare our method with the variant REALISM w/o R, as shown in Table~\ref{tab:performance}. The results show that without regulations, operators focus only on maximizing their own profits, leading to poorer city goal achievement, which is similar to that of the model RECOMMEND. This underscores the importance of the city regulation component in achieving city goals.

\subsubsection{\textbf{The Effectiveness of Fairness Awareness in Score Assignment}} To demonstrate the effectiveness of fairness awareness in score assignment, we conduct comparison experiments with the variant REALISM w/o FASA, as shown in Table~\ref{tab:performance}. The results indicate that without fairness awareness in score assignment, the city regulator's regulations can still promote better achievement of city goals compared to the baselines.
We use a simple example to explain the reason: If achieving the city goal requires 100$\%$ effort, fairness only plays a role in how to distribute these efforts among the three operators based on their net revenue or fleet size. This distribution may not necessarily affect the achievement of the city goal. 
However, those regulations without fairness awareness do so by providing unfair feedback, resulting in operators with higher profits sacrificing more profits for cooperation. On the contrary, operators with lower profits sacrifice less profit. Therefore, the findings in Table~\ref{tab:performance} indicate the issue of disparity in operators’ contributions that even though the performance of REALISM w/o FASA in city goal achievement is close to REALISM, higher-earning operators (Spin and Bird) contribute more to city goals, whereas lower-earning ones (Lime) contribute less. This in turn, validates the importance of considering fairness in scoring.

\subsubsection{\textbf{The Impact of Different Fairness Definitions}}
The definition of fairness differs in different situations~\cite{saxena2019fairness}. 
Here we introduce a new fairness definition to evaluate how it impacts the model performance.
Following~\cite{jiang2023faircod}. To define vehicle usage equity, we use the difference between the regional demand satisfaction rate and the average demand satisfaction rate across all regions, aiming to achieve equal satisfaction rates across all regions. To define score assignment fairness, we use the difference between operator scores and average operator scores, aiming to achieve equal scores across all the operators. They are respectively formulated as $C^{equ}_{t}=-\frac{1}{N}\sum_{i=1}^{N}|\frac{D^{i}_{t}}{F^{i}_{t}}-\frac{\sum^{N}_{i=1}D^{i}_{t}}{\sum^{N}_{i=1}F^{i}_{t}}|$,
$E_{t}=-\sum^{M}_{m=1}|Z^{pen,m}_{t}-\frac{\sum^{M}_{m=1}Z^{pen,m}_{t}}{M}|$. The notation meanings are detailed in Section~\ref{sec:problem}.

\begin{figure}[h]\centering
\vspace{-10pt}
\begin{minipage}[t]{0.45\linewidth}
    \includegraphics[width=\linewidth, keepaspectratio=true]{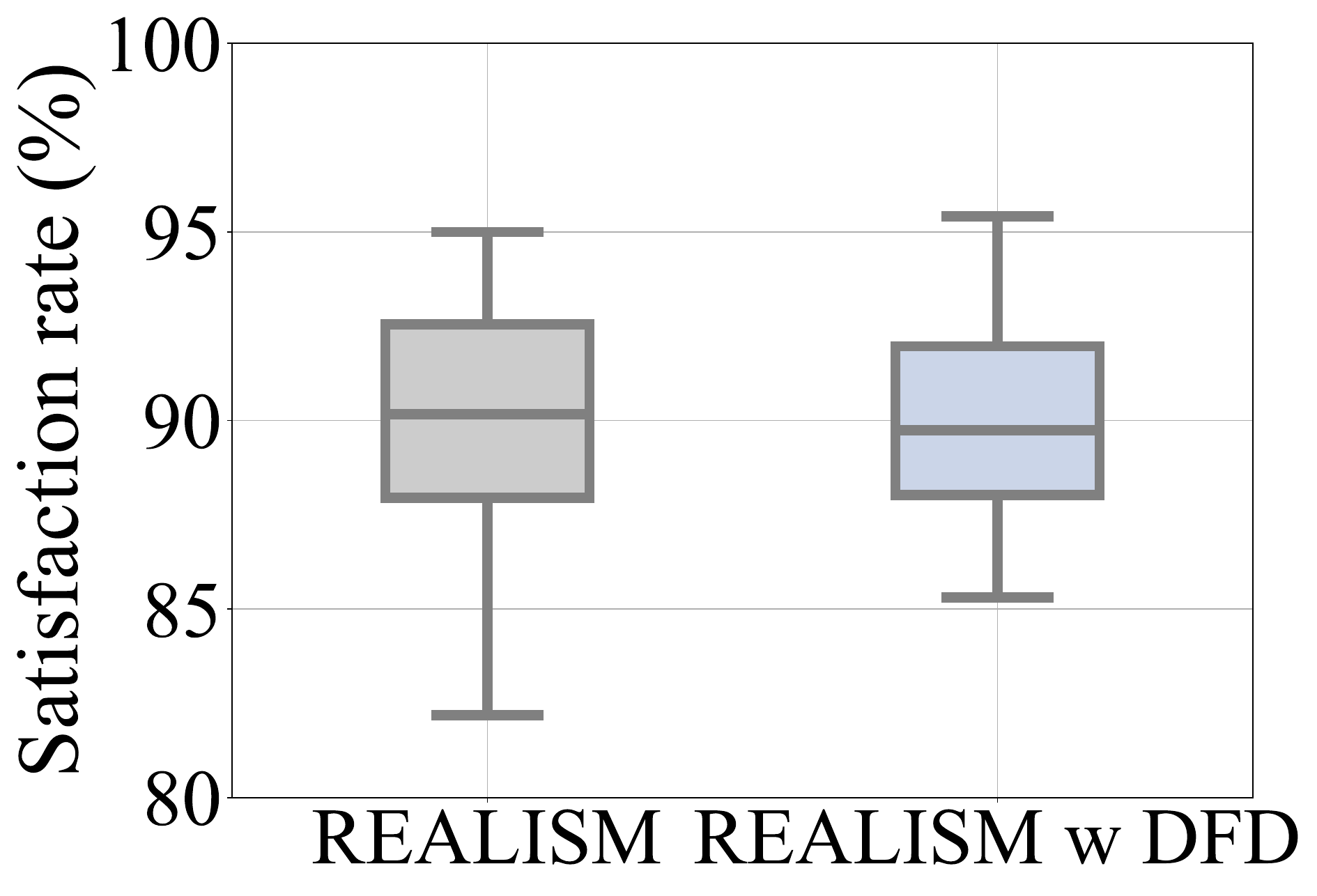}
    \vspace{-15pt}
    \captionsetup{font={small}}
    \caption{The regional distribution of demand satisfaction}
    \label{fig:fair_sat}
\end{minipage}
\hspace{10pt}
\begin{minipage}[t]{0.47\linewidth}
    \includegraphics[width=\linewidth, keepaspectratio=true]{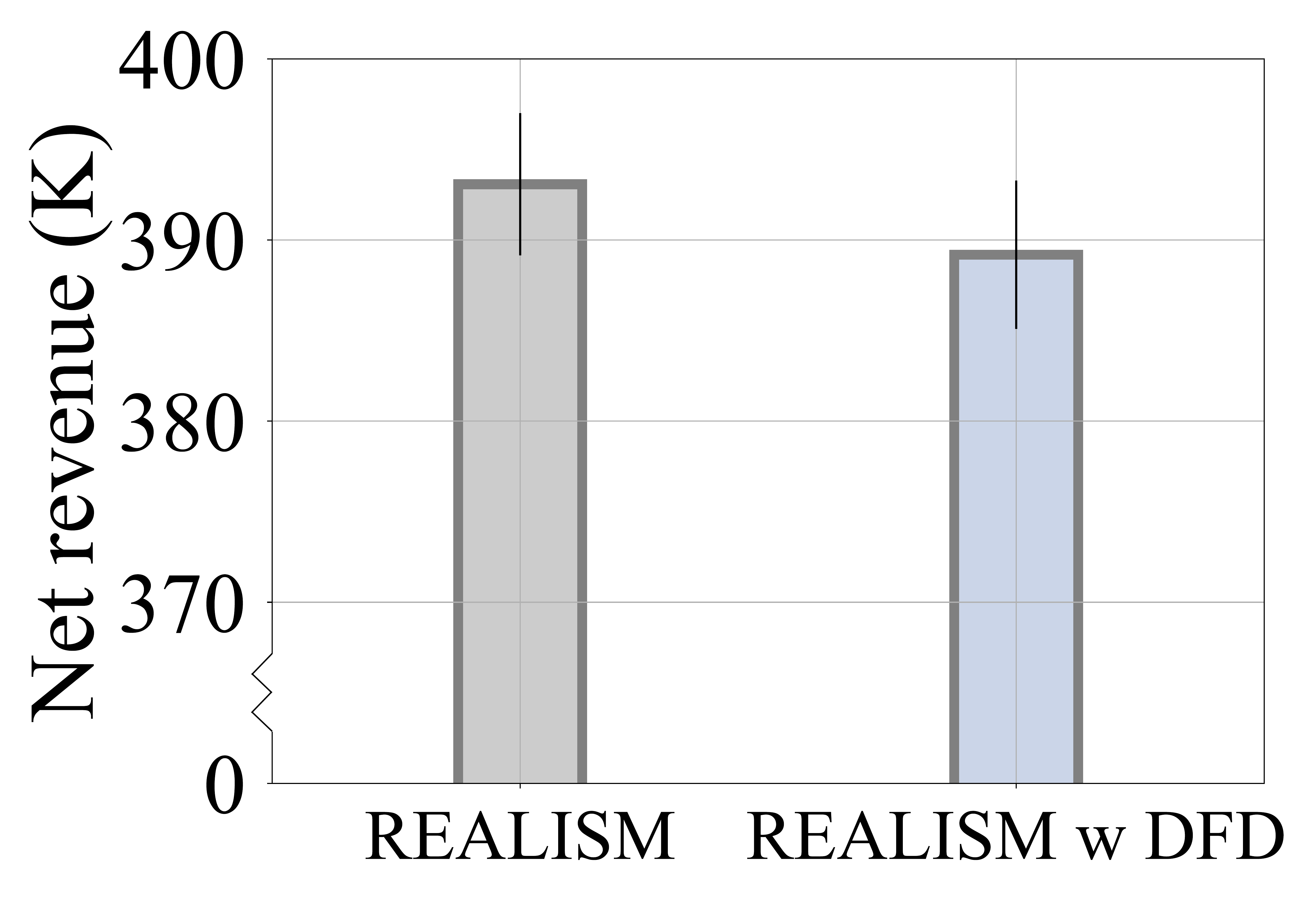}
    \vspace{-15pt}
    \captionsetup{font={small}}
    \caption{The monthly average net revenue of operators}
    \label{fig:fair_rev}
\end{minipage}
\vspace{-10pt}
\end{figure}

Figure~\ref{fig:fair_sat} and Figure~\ref{fig:fair_rev} show the results of comparison experiments with our model variant (i.e., REALISM w DFD) in terms of satisfaction rates and net revenue. Specifically, each box in Figure~\ref{fig:fair_sat} represents the distribution of demand satisfaction rates across all city regions, aggregated across three operators. The net revenue in Figure~\ref{fig:fair_rev} is the average of the three operators.
From Figure~\ref{fig:fair_sat}, we know that the average regional demand satisfaction rates of the two models are nearly the same while REALISM w DFD achieves a smaller fluctuation. The possible reason is that REALISM w DFD aims to achieve equal satisfaction rates across all regions according to its definition of vehicle usage equity and both REALISM and REALISM w DFD share the same city goal of maximizing the demand satisfaction rate. 
However, Figure~\ref{fig:fair_rev} reveals that REALISM achieves higher net revenue for operators in the city compared to REALISM w DFD. The possible reason is that the focus on balancing regional demand satisfaction may undermine the operators' profits.
In addition, balancing the amount of scores without accounting for operators' net revenue could further reduce their profits. 

\vspace{-10pt}
\begin{figure}[h]\centering
\begin{minipage}[h]{0.45\linewidth}
    \includegraphics[width=\linewidth, keepaspectratio=true]{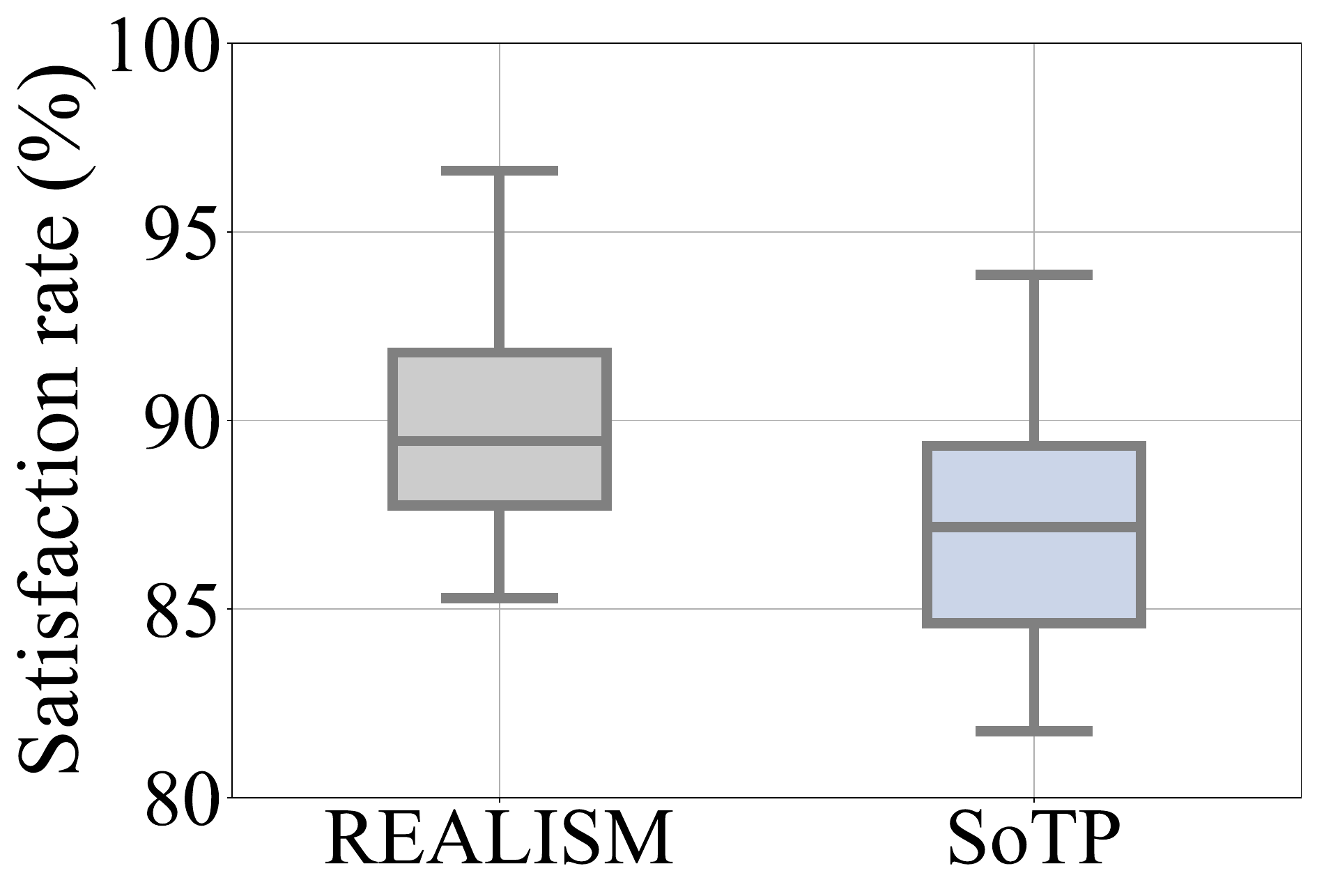}
    \vspace{-15pt}
    \captionsetup{font={small}}
    \caption{The distribution of demand satisfaction rate in one month}
    \label{fig:case_sat}
\end{minipage}
\hspace{10pt}
\begin{minipage}[h]{0.45\linewidth}
    \includegraphics[width=\linewidth, keepaspectratio=true]{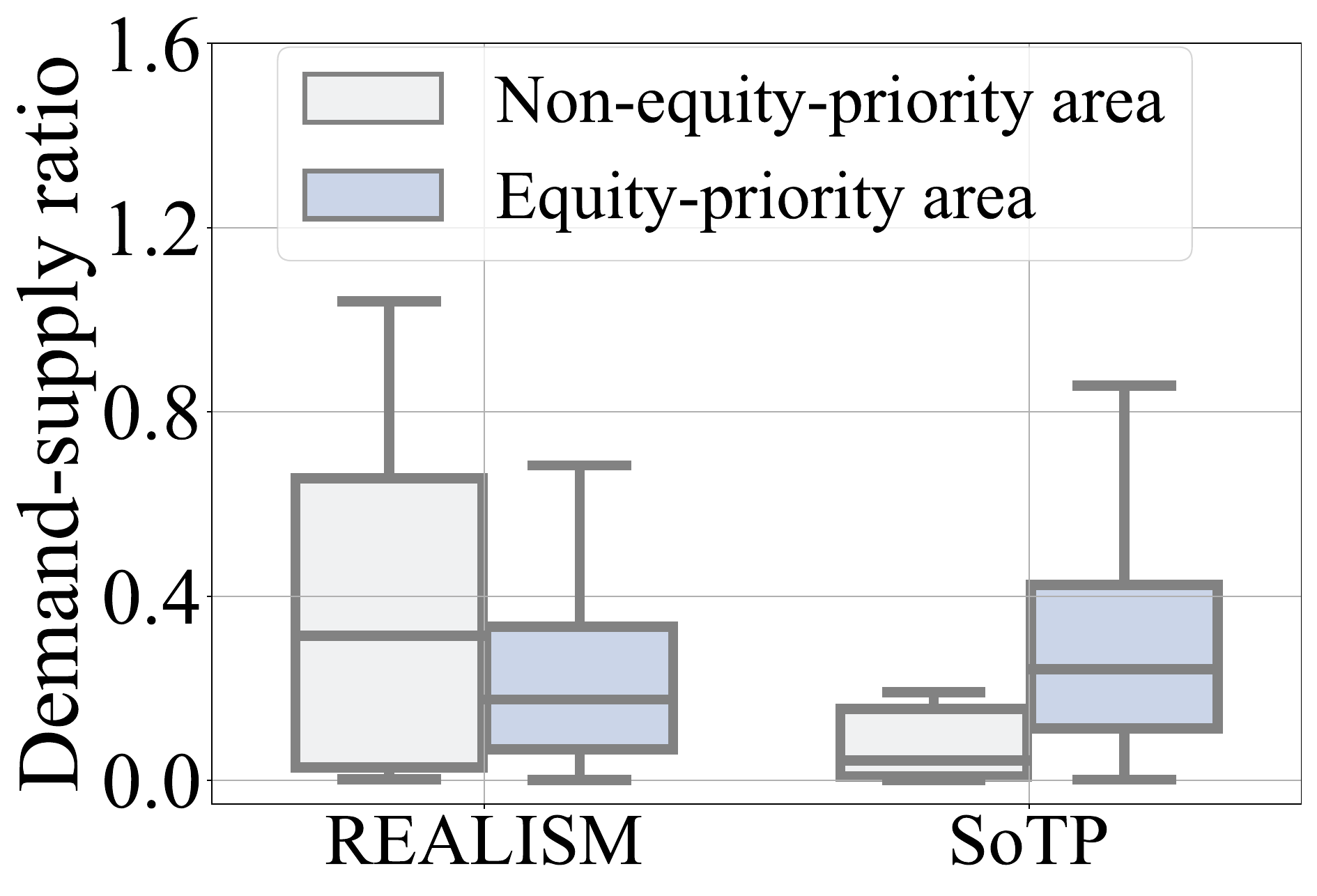}
    \vspace{-15pt}
    \captionsetup{font={small}}
    \caption{The distribution of vehicle usage equity in one month}
    \label{fig:case_equ}
\end{minipage}
\vspace{-10pt}
\end{figure}

\subsection{Case Study}

We present a case study to demonstrate the effectiveness of REALISM in achieving city goals related to both the demand satisfaction rate and vehicle usage equity compared to the state-of-the-practice regulation (SoTP) used in Chicago~\cite{chicago_rule}.
Specifically, SoTP conducts strict regulation by forcing a minimum number of vehicles in certain areas (i.e., 62 equity-priority regions in Chicago covering 42.2\% of the areas).
Figure~\ref{fig:case_sat} illustrates the distribution of the daily demand satisfaction rate in one month. 
It shows that compared with SoTP, the flexible regulations of REALISM enable operators to achieve a higher demand satisfaction rate.
Figure~\ref{fig:case_equ} displays the distribution of vehicle usage equity in two categories of areas (i.e., equity priority areas and non-equity priority areas).
We found that our method achieves a comparable good result in the equity-priority areas, without losing the flexibility of rebalancing vehicles, compared to the strict number of vehicles in these regions using SoTP.
In non-equity priority areas, our method achieves a better demand-supply ratio (i.e., close to one), which satisfies the most demand while resulting in fewer idle vehicles. This case study demonstrates the practical benefits that our method can offer to cities.


\section{Discussion}
\noindent \textbf{Lessons}: Based on the results from our work, we summarize the following learned lessons:
(i) The city regulator's regulations make operators aware of the achievement of city goals and collaboratively realize them, as shown in Table~\ref{tab:performance}. Without regulations, operators only focus on maximizing their own profits, leading to worse city goal achievements.
(ii) The city regulator's regulations, which assign scores, should ensure fairness of the assigned scores across all operators, as demonstrated in Table~\ref{tab:performance}. Lack of fairness consideration can result in operators receiving unfair scores and consequently making unfair contributions to the city goal achievements.

\noindent \textbf{Limitations}: (i) We assume that the city regulator regulates operators only in the form of score assignment. The flexible forms of regulations on multi-operator vehicle rebalancing will be our future work.
(ii) We utilize a single dataset for evaluation, as the dataset from Chicago is the only publicly available resource that offers spatial-temporally synchronized multi-operator data. We plan to extend our evaluation by incorporating additional datasets from multiple cities in future work. (iii) In our paper, we aim to develop a general regulatory framework for multi-operator shared micromobility vehicle rebalancing in a city: the city regulator regulates the shared micromobility operators’ vehicle rebalancing in a city and gives feedback to each operator considering different perspectives, including demand satisfaction, vehicle usage equity, total ridership, etc. The feedback can incentivize or punish those operators to guide them to adjust their rebalancing strategies, and the feedback sent to each operator should be fairness-aware. Therefore, the metrics of vehicle usage equity and fairness of score assignment can be replaced by other works (e.g., worst-case notions of fairness~\cite{diana2021minimax}, and causal notions of fairness~\cite{kusner2017counterfactual}).
(iv) In our work, we assume a non-competitive relationship between operators that users rely on a single operator—they do not switch between operators even when their preferred operator has insufficient supply. Under the assumption that users utilize services from multiple operators, the operators’ relationship would shift to a competitive one, thereby imposing greater complexity and challenges for the regulatory framework. This is an interesting direction to explore in the future.

\noindent \textbf{Incentive mechanism}: Even though it is ideal that the city regulator regulates operators' vehicle rebalancing in the form of incentivizing their contributions to the city goal achievement, designing a universal incentive mechanism is tough due to the challenge of measuring operators' responses to incentives, such as their acceptance and preferences. Our framework allows regulators to influence operators' rebalancing strategies and automate strategy adjustments by incorporating monetary scores into the reinforcement learning process. Practically, rewards for meeting city goals could include fleet expansion, whereas failures may lead to license suspensions \cite{chicago_rule}. These outcomes could be translated into scores by carefully assessing their monetary equivalent.

\noindent \textbf{Model Generalizability}: While the experimental setup is based on Chicago’s 77 regions and 3 operators, our framework is designed to be general and adaptable across a broad range of urban mobility settings: (i) The vehicle scheduling module in our framework is modular and can be replaced by different baseline algorithms, including those that handle heterogeneous fleets such as bikes and scooters with different operational constraints and demand dynamics. (ii) The metrics used by the regulator in the score assignment mechanism to evaluate each operator’s contribution toward city goal achievement are also replaceable. In addition, those metrics are modular and can be tailored to reflect local regulations, geographic contexts, or the specific characteristics of scheduled vehicle types in different cities. (iii) Our framework supports dynamic adjustment in the number of operators, and its underlying mechanisms remain effective regardless of the scale.


\section{Related Work}
Multi-fleet management aims to achieve cost savings and superior service quality by efficiently managing a variety of vehicle fleets.
Existing works can be divided into two categories, including (i) cooperative-based multi-fleet management and (ii) non-cooperative-based multi-fleet management.

In the cooperative paradigm, one solution involves establishing multiple agents to control the fleet. Some
studies assume a unified objective among all fleets, collectively striving towards its attainment~\cite{tan2023joint,tan2023joint1,zhang2022multi,pan2019deep,kondor2022cost,sun2024optimizing,li2019efficient,zhang2021intelligent,oroojlooy2023review,yang2024mallight}.
For example, RECOMMEND~\cite{tan2023joint} views each region as an agent responsible for rebalancing and charging shared electric micromobility vehicles. These agents share information and are dedicated to a common goal: finding the optimal charging and rebalancing strategies that maximize the platform's benefits. 
MAGC~\cite{zhang2022multi} aims to solve the charging station request-specific dynamic pricing problem. They regard each charging station as an agent and consider the complex, competitive, and cooperative relationships between different agents to maximize operator benefits through vehicle scheduling.
However, in practice, this assumption often falters since operators have divergent objectives driven by their individual profit motives, which render these cooperative approaches less applicable.
Another solution is to establish a central controller to facilitate collaboration. Kondor~\textit{et al.}~\cite{kondor2022cost} control agents in an intrusive way by directly influencing their decision-making.
However, it may not be applicable for our work as operators might not fully adhere to directives from such a centralized entity and tend to retain control of their fleets. In addition, the setting of the single control agent may introduce a large action and state space, which is computationally intractable.

In the non-cooperative-based methods, they establish multiple agents to manage the fleet, with each agent pursuing its own objectives and striving to maximize its individual goals~\cite{wang2021record,li2021dynamic,lin2018efficient,wang2020joint,yan2024robust,zhao2024urban}.
For example, Record~\cite{wang2021record} designs a dynamic deadline-based distributed deep reinforcement learning algorithm to learn the most suitable service station for relocation and the optimal charging station for each shared electric vehicle. They consider each unoccupied shared electric vehicle as an agent and enable it to maximize its own objectives.
STMIP~\cite{li2021dynamic} considers the shared bike rebalancing problem as a truck-based rebalancing problem and models it as a spatial-temporal mixed integer programming problem. 
However, due to the absence of central regulation, their frameworks make agents unable to communicate with each other and achieve the collaborative goal. 

\section{Conclusion}
In this work, we focus on the problem of multi-operator vehicle scheduling under the regulation set by the city regulator. We design a MARL-based framework called REALISM, which incorporates a fairness-aware score assignment prediction model to regulate operators' vehicle rebalancing. In REALISM, we utilize an alternating procedure to optimize both the vehicle rebalancing agents and the score assignment prediction model. The evaluation results demonstrate that REALISM achieves an improvement of at least 39.93\% in vehicle usage equity and 1.82\% in the average demand satisfaction across the whole city, compared to state-of-the-art baselines.

\section{Acknowledgement}
This work was supported in part by NSF grants 2246080, 2318697, and 2427915. Yukun Yuan was partially supported by NSF grant 2431552, and Guang Wang was partially supported by NSF grant 2411152 and the FSU FYAP award.


\bibliographystyle{ACM-Reference-Format}
\bibliography{refs}

\end{document}